\documentclass{cpbtex}
\usepackage{graphicx}
\usepackage{amssymb}
\usepackage{amsmath}
\usepackage{mathrsfs}
\usepackage{color}
\usepackage{mathtools}
\usepackage{physics}
\usepackage{bm}
\usepackage{siunitx}
\usepackage{lineno}
\usepackage[final]{changes}
\usepackage{soul}
\begin{document}
% \begin{CJK*}{GBK}{song}

\title{Threshold-independent method for single-shot readout of spin qubits in semiconductor quantum dots}

\author{Rui-Zi Hu$^{1,2}$, Sheng-Kai Zhu$^{1,2}$, Xin Zhang$^{1,2}$, Yuan Zhou$^{1,2}$, Ming Ni$^{1,2}$, \\ Rong-Long Ma$^{1,2}$, Gang Luo$^{1,2}$, Zhen-Zhen Kong$^{3}$, Gui-Lei Wang$^{3,4}$, Gang Cao$^{1,2}$, \\ Hai-Ou Li$^{1,2}$\thanks{E-mail: haiouli@ustc.edu.cn}, Guo-Ping Guo$^{1,2,5}$\\
$^{1}${CAS Key Laboratory of Quantum Information,} \\ {University of Science and Technology of China, Hefei, Anhui 230026, China}\\  
$^{2}${CAS Center For Excellence in Quantum Information and Quantum Physics,} \\ {University of Science and Technology of China, Hefei 230026, China}\\ 
$^{3}${Key Laboratory of Microelectronics Devices and Integrated Technology,} \\{Institute of Microelectronics, Chinese Academy of Sciences, Beijing 100029, China}\\
$^{4}${Beijing Superstring Academy of Memory Technology, Beijing 100176, China}\\
$^{5}${Origin Quantum Computing Company Limited, Hefei, Anhui 230026, China}} 
% The line break was forced via \\

\date{\today}
\maketitle

\begin{abstract}
    The single-shot readout data process is essential \highlight{for the 
    realization of high-fidelity qubits and} fault-tolerant quantum algorithms 
    in semiconductor quantum dots. However, the fidelity and visibility of the 
    readout process is sensitive to the choice of the thresholds and limited by 
    the experimental hardware. By demonstrating \highlight{the linear 
    dependence between} the measured spin state probabilities and readout 
    visibilities 
    \highlight{along with} dark counts, we describe an alternative 
    threshold-independent method for the single-shot readout of spin qubits in 
    semiconductor quantum dots. We can obtain \highlight{the extrapolated spin 
    state probabilities of the prepared probabilities of the excited spin 
    state} through the threshold-independent method. Then, we analyze the 
    corresponding errors of the method, \highlight{finding that} errors of the 
    extrapolated probabilities cannot be neglected with no constraints on the 
    readout time and threshold voltage. \highlight{Therefore, by limiting the 
    readout 
    time and threshold voltage we ensure the accuracy of the extrapolated 
    probability.} Then, we \highlight{prove that the efficiency and robustness 
    of this 
    method is 60 times larger than that of the most commonly used method.} 
    Moreover, we discuss the influence of the electron temperature on the 
    effective area with a fixed external magnetic field and provide a 
    preliminary demonstration for a single-shot readout up to 0.7 K/1.5 T in 
    the future.
\end{abstract}



\section{\label{sec:Intro}Introduction}

Spin qubits in gate-defined silicon quantum dots (QDs) are promising for 
realizing quantum computation due to their long coherence 
time~\cite{Muhonen2014Storing,Zhang2018Semiconductor}, small 
footprint~\cite{Eriksson2020fabtication}, potential 
scalability~\cite{Veldhorst2018scalable}, and industrial 
manufacturability~\cite{Camenzind2021Spin,Zwerver2022Qubits}. In isotopically 
purified Si devices, the single-qubit gate fidelity has attained 
99.9\%~\cite{Tarucha2018single-qubit, Dzurak2018single-qubit}, and the 
two-qubit gate fidelity above 99\% has been 
reported~\cite{Xue2022Qunatum,Noiri2022Fast,Mills2022Tow-qubit}. However, the 
corresponding readout fidelities are lower than 99$\%$\highlight{, 
significantly reducing the overall fidelity of gate operation.} 
\highlight{We usually use the Elzerman single-shot readout 
method}~\cite{elzerman2004single,Zajac2016Scable}, which utilizes the 
spin-state-dependent tunneling rate or quantum capacitance to measure the spin 
qubits with extra charge sensors or dispersive sensing 
techniques~\cite{Pakkiam2018single-shot,West2019Gate-based,Zheng2019RapidReadout}.
 The spin states are distinguished by comparing the readout traces $x$ with the 
threshold voltages $x_{\rm t}$ within the readout time $t_{\rm r}$. However, 
this process is sensitive to $x_{\rm t}$ and $t_{\rm r}$, and will lower the 
overall fidelity of the gate operations.

To optimize the readout fidelity $F^R$ and visibility $V^R$, \highlight{as well 
as to} find out the corresponding optimal threshold voltage $x_{\rm t}$ and 
readout time $t_{\rm r}$, \highlight{we have tried several methods}, e.g., 
wavelet edge detection~\cite{Prance2015identifying}, the analytical expression 
of the distribution~\cite{Nowack2011single-shot,anjou2014optimal}, statistical 
techniques~\cite{gorman2017tunneling}, neural network~\cite{Struck2021Neural}, 
digital processing~\cite{Mizokuchi2020Detection} and the Monte-Carlo 
method~\cite{merello2010single}. Among these \highlight{different methods}, the 
Monte-Carlo method is now widely used to numerically simulate the distributions 
of the experimental data in Si-MOS QDs~\cite{veldhorst2014addressable}, Si/SiGe 
QDs~\cite{Kawakami2014electrical}, Ge QDs~\cite{Vukusic2018single-shot}, single 
donors~\cite{Buch2013spin}, and nitrogen-vacancy 
centers~\cite{robledo2011high-fidelity}. A high fidelity readout in a silicon 
single spin qubit has been achieved recently~\cite{Mills2022High_Fidelity}. 
Even so, the readout visibility $V^R$ is limited by the environment and 
experimental setup, e.g., the external magnetic field relative to the electron 
temperature ($B_{\rm ext}/T_{\rm e}$), relaxation time ($T_1$), tunneling rate 
($\Gamma^{\rm in,out}$), measurement bandwidth, sample rate($\Gamma_{\rm s}$), 
and filter frequency~\cite{keith2019benchmarking}.

Here, we describe a threshold-independent method for the single-shot readout of 
semiconductor spin qubits. By \highlight{considering} the rate 
equations~\cite{keith2019benchmarking,xiao2010measurement,Carroll2013electron,house2013detection}
 \highlight{and} the Monte-Carlo method, we simulate the single-shot readout 
process and extract $V^R$ as a function of readout time $t_{\rm r}$ and 
threshold voltage $x_{\rm t}$. We demonstrate that the measured probability of 
the excited spin state ($P^{M}_\uparrow$) is linearly dependent on $V^R$ 
\highlight{in} Eq.~\ref{eq: relation}. Since the slope is the prepared 
probability of the excited spin state ($P^I_\uparrow$) and is robust to $t_{\rm 
r}$ and $x_{\rm t}$, \highlight{it is convenient to} use $P^I_\uparrow$ instead 
of $P^{M}_\uparrow$ to realize a threshold-independent data processing method. 
Then, we analyzed the error of the fitting process, \highlight{finding} that 
the error from the bin edges \highlight{caused to a discrepancy between the 
result and the expected value.} \highlight{We ensured accurate extrapolated 
probability by choosing} readout time $t_{\rm r}$ and threshold voltage $x_{\rm 
t}$. Moreover, we use an effective area ($A_{\rm eff}$) to show that the 
effectiveness of the threshold-independent method. \highlight{It's 
approximately 60 times larger than the commonly used method}, i.e., the 
threshold-dependent method. Finally, we discussed the influence of $T_{\rm e}$ 
on the effective area of both the threshold-independent method and the 
threshold-dependent method with a fixed external magnetic field 
\highlight{along with providing} a preliminary demonstration for a single-shot 
readout at 0.7 K/1.5 T in the future.

\section{\label{sec:Result and Discussion}Results and Discussion}

\subsection{\label{sec:Single-shot}Single-shot readout}

\begin{figure}[t]
  \centering 
  \includegraphics[width=8.6cm]{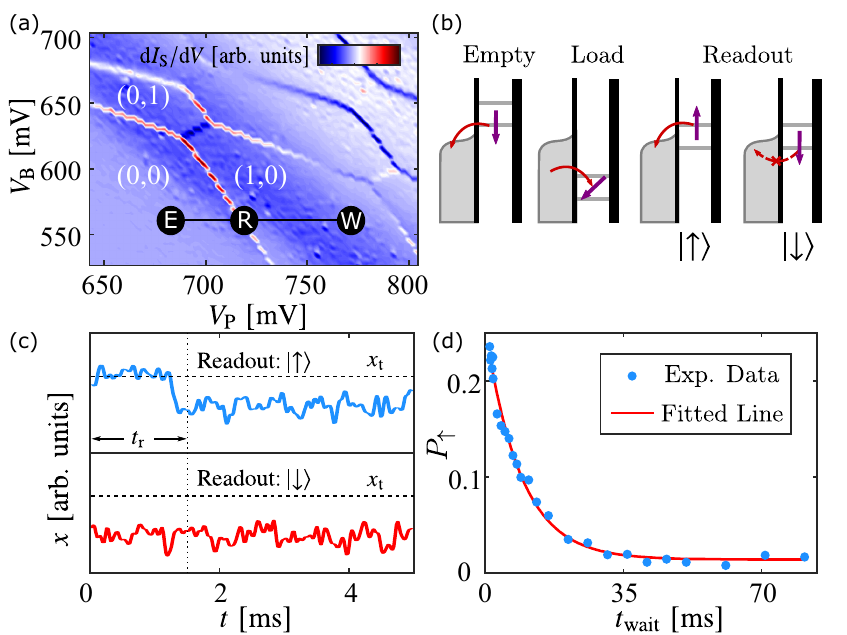}
  \caption{\label{fig:1}(a) Charge stability diagram of the DQD measured by 
  differentiating the single-electron transistor (SET) current $(I_S)$ as a 
  function of the $V_B$ and $V_P$ gate voltages. The pulse sequence for 
  measuring the spin relaxation time ($T_1$) via the (0,0)-(1,0) charge 
  transition line is overlaid on the data. (b)	Illustration of the energy 
  states of the pulse sequence: the measurement starts with emptying the 
  electron in the QD at point E, then injecting a random spin into it and 
  waiting for a time at point W, finally moving to point R for STC conversion. 
  (c) The readout traces ($x$) are achieved by amplifying $I_S$ with a room 
  temperature low noise current amplifier (DLCPA - 200) and a JFET preamplifier 
  (SIM910), and then low-pass filtering the amplified signal using an analog  
  filter (SIM965) with a bandwidth of 10 kHz. The blue and red curves represent 
  spin up ($\ket{\uparrow}$) and spin down ($\ket{\downarrow}$) traces at point 
  R, respectively. By comparing the maximum of each readout traces ($x_{\rm 
  max}$) with the threshold voltage ($x_{\rm t}$) within the readout time 
  ($t_{\rm r}$), we can distinguish the different spin states. (d) A typical 
  exponential decay of spin-up probability with \highlight{1000} repeated 
  measurements for each point. The exponential fit \highlight{for} $1/T_1$ is 
  $112\pm6$ s$^{-1}$}.
\end{figure}

Fig.~\ref{fig:1} outlines the processes of single-shot readout. The double 
quantum dots (DQDs) in our experiment \highlight{resemble} the device in 
Ref.~\cite{Zhang2020Giant}. ($N_L,\ N_R$) in the charge stability diagram in 
Fig.~\ref{fig:1}(a) represent the electrons occupied in the left and right QD. 
To measure the spin state of the first electron in the left QD, we deploy 
consecutive three-stage pulses, which consist of “empty”, “load \& wait” and 
“readout” at (0,0)-(1,0) transition line, illustrated by “E”, “W” and “R” in 
black circles in Fig.~\ref{fig:1}(a). Fig.~\ref{fig:1}(b) shows the 
corresponding energy states. Here, we assume that the excited spin state is 
$\ket{\uparrow}$. The location of the readout stage is carefully calibrated to 
ensure that the Fermi level of the reservoir is between the electrochemical 
potentials of spin-up and spin-down states. 

The readout traces \highlight{we measured} by amplifying the single-electron 
transistor (SET) current ($I_S$) with a room temperature low noise current 
amplifier (DLCPA - 200) and a JFET preamplifier (SIM910) and then low-pass 
filtering the amplified signal using an analog  filter (SIM965) with a 
bandwidth of 10 kHz. \highlight{The blue and red curve represent spin up and 
spin down traces measured at point R, respectively. By comparing the maximum 
value of each readout traces with the threshold voltage within the readout 
time, we can distinguish the different spin states.}

The state-to-charge (STC) conversion is realized by distinguishing two 
different traces in the readout phase, as shown in Fig.~\ref{fig:1}(c). Single 
electron tunneling onto or off the QD causes a change in the readout trace $x$. 
We distinguish the different spin states in the QD by comparing readout trace 
$x$ with the threshold voltage $x_{\rm t}$. If $x$ remains below the threshold 
during the readout phase, we assume it is a $\ket{\downarrow}$ state and vice 
versa. Then, we \highlight{emptied} QDs by raising the electrochemical 
potential and waiting for enough time. After the empty stage, we 
\highlight{loaded} a new electron with a random spin state and 
\highlight{waited} for the next readout stage. 

We \highlight{measured} the electron spin relaxation time by repeating this 
three-stage pulse and changing the waiting time in the loading stage. 
Fig.~\ref{fig:1}(d) shows \highlight{typical} exponential decay of the measured 
spin-up probability  $P^M_\uparrow = \rho\cdot e^{-t/T_1}+\alpha$, where $\rho$ 
is the amplitude and $\alpha$ is the dark count. Additionally, we can 
manipulate the spin qubit by using a similar pulse and a microwave pulse, as 
reported in Ref.~\cite{Hu2021Operation}.

\subsection{\label{sec:readout Visibility}The readout visibility}
\begin{figure}[t]
  \centering  
  \includegraphics[width=8.6cm]{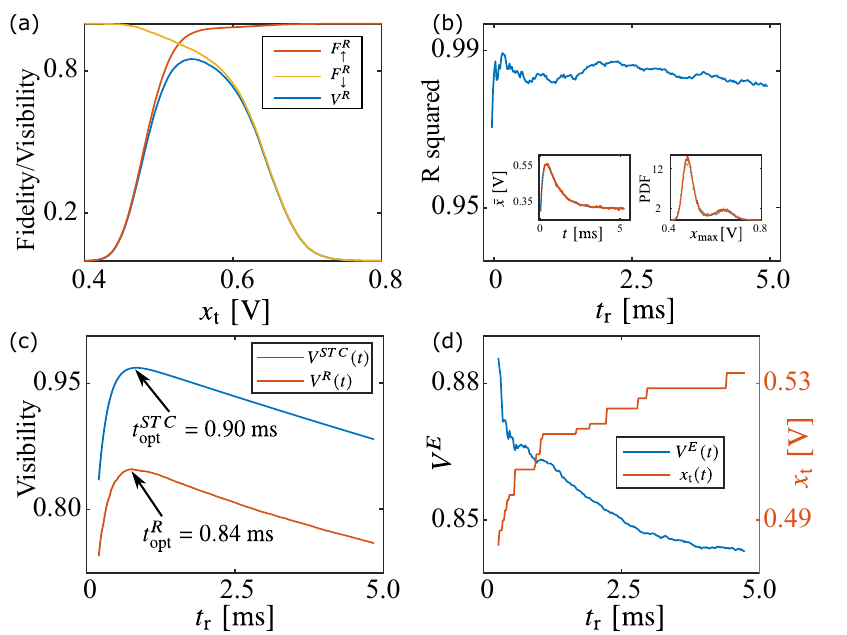}% Here is how to import EPS art
  \caption{\label{fig:2} (a) The fidelities of spin-up and spin-down state 
  ($F_\uparrow^R, F_\downarrow^R$) and the related readout visibility ($V^R$) 
  versus $x_{\rm t}$. (b) The fitting model for the distribution of the 
  experimental data has a high value of R squared ($\approx 0.98$), which 
  indicates good accuracy of the results. The insets show fitting results for 
  the averaged traces ($\bar{x}$) and the probability density function (PDF) of 
  the maximum \highlight{value} of each readout traces ($x_{\rm max}$). (c) The 
  state-to-charge (STC) conversion visibility ($V^{STC}$) versus $t_{\rm r}$ 
  and the corresponding maximum \highlight{value} $V^R$. $t_{\rm opt}^R$ (the 
  bottom arrow) and $t_{\rm opt}^{STC}$ (the top arrow) are not equal. (d) The 
  electrical detection visibility ($V^E$) and the corresponding optimal 
  threshold voltage $x_{\rm t}$ as a function of readout time $t_{\rm r}$.
}
\end{figure}

We \highlight{demonstrated} a maximum visibility $V^R=85.4\%$ while measuring 
the spin relaxation time as shown in Fig.~\ref{fig:2}(a). 
$V^R=F_\uparrow^R+F_\downarrow^R-1$ is calculated from the simulated data via 
the Monte-Carlo method. $F_\downarrow^R$ and $F_\uparrow^R$ are the readout 
fidelities of $\ket{\downarrow}$ and $\ket{\uparrow}$. In Fig.~\ref{fig:2}(b), 
\highlight{we simulated the distribution of the single-shot signal with a high 
accuracy (R squared $\approx0.98$) of the corresponding fitting process} (the 
right inset in Fig~\ref{fig:2}.(b)). The insets in Fig~\ref{fig:2}.(b) shows 
the fitting results of the averaged readout traces ($\bar{x}$) and the 
probability density function (PDF) of the traces maximum ($x_{\rm max}$) of the 
readout phase for every single measurement. The details about the simulation 
process and the insets in Fig.~\ref{fig:2}(b) are discussed in Sec.~1 of the 
Supplementary Materials~\cite{merello2010single, 
Buch2013spin,keith2019benchmarking,xiao2010measurement, Carroll2013electron}. 
By \highlight{fitting $\bar{x}$ with the rate equations}, we 
\highlight{obtained} the tunneling rates of the state-to-charge conversion as 
shown in the left inset of Fig.~\ref{fig:2}(b): $\Gamma_{\uparrow}^{\rm out} = 
6.0\pm0.1\ {\rm kHz}$, $\Gamma_{\downarrow}^{\rm out}=27\pm2\ {\rm Hz}$ and 
$\Gamma_\downarrow^{\rm in}=1.39\pm0.04\ {\rm kHz}$.

In the traditional \highlight{methods}, state-to-charge conversion visibility 
($V^{STC}$) and electrical detection visibility ($V^E$) need to be optimized to 
obtain a high readout visibility $V^R$. \highlight{Instead, the 
threshold-independent single-shot readout method does not need to perform such 
a cumbersome operation. Here we introduce the steps of this novel method.} 
First, we \highlight{calculated} $V^{STC}$ as a single peaked function of 
readout time $t_{\rm r}$ with the knowledge of tunneling rates as shown in 
Fig.~\ref{fig:2}(c). (We \highlight{will} describe the details of this part in 
Sec.~2 of the Supplementary Materials.) For exceeding 99\% $V^{STC}$, three 
criteria are given in Ref.~\cite{keith2019benchmarking}, including $E_z/T_{\rm 
e}>13$, $T_1*\Gamma^{\rm out}_\uparrow>100$ and $\Gamma_{\rm s}/\Gamma^{\rm 
in}_\downarrow>12$. Here, we have $\Gamma_{\rm s}=50$ kHz, $B_{\rm ext} = 1.5$ 
T and $T_{\rm e}=180.5\pm8.1$ mK as mentioned in Ref.~\cite{Zhang2020Giant}; 
thus, the disagreement of the condition $E_z/T_{\rm e}=11.22<13$ limits the 
maximum $V^{STC}$ to 97.15\%. Fig.~\ref{fig:2}(c) also shows the readout 
visibility $V^{R}$ as a single peaked function of $t_{\rm r}$ (details are 
discussed in Sec. 4 of the Supplementary Materials). However, the optimal 
readout time $t_{\rm opt}$ for $V^{STC}$ is different \highlight{from} that for 
$V^{R}$. Thus, \highlight{to obtain the maximum $V^R$, we should consider} the 
readout time $t_{\rm r}$ and threshold voltage $x_{\rm t}$ together.
% It reveals that readout time $t_{\rm r}$ and threshold voltage $x_{\rm t}$ are not independent while optimizing readout visibility $V^R$, and they should be considered simultaneously to obtain the maximum $V^R$ instead of being deemed in order.

Due to the lack of a simple analytical expression for the distribution of $x_{\rm max}$, $V^E$ cannot be obtained directly. Here, we obtain $V^E$ by factorizing $V^R$ into $V^{STC}=F_\uparrow^{STC}+F_\downarrow^{STC}-1$ and $V^E=F_\uparrow^E+F_\downarrow^E-1$, as the following formula shows:
\begin{eqnarray}
  V^{R} =F_\uparrow^R+F_\downarrow^R-1=V^{STC}\times V^{E}.
  \label{eq:readout visibility}
\end{eqnarray}
Here, readout fidelities $F_\downarrow^R$ and $F_\uparrow^R$ are 
\highlight{shown} the following:
\begin{eqnarray}
  F_\downarrow^R &&= F_\downarrow^{STC} F_\downarrow^E + (1-F_\downarrow^{STC})(1-F_\uparrow^E),\nonumber\\
  F_\uparrow^R &&= F_\uparrow^{STC} F_\uparrow^E + (1-F_\uparrow^{STC})(1-F_\downarrow^E),
  \label{eq:readout fidelity}
\end{eqnarray}
$t_{\rm r}$, the corresponding readout visibility $V^R$ is obtained from the 
Monte-Carlo method and the state-to-charge visibility $V^{STC}$ is calculated 
as above. Thus, we have the electrical detection visibility $V^E=V^R/V^{STC}$ 
indirectly. Fig.~\ref{fig:2}(d) shows the maximum $V^E$ and the corresponding 
optimal threshold voltage $x_{\rm t}$ \highlight{versus} readout time $t_{\rm 
r}$. \highlight{Considering the increasing or constant nature of the maximum 
value $x_{\rm t}$ as $t_{\rm r}$ increases in each readout trace, it can be 
inferred that the optimal threshold voltage $x_{\rm t}$ follows a monotonically 
non-decreasing pattern. On the other hand, the longer $t_{\rm r}$ we consider, 
the more noise is added to the trace. Therefore, it is obvious that $V^E$ will 
decrease as $t_{\rm r}$ increases.}
% \section{\label{sec:3}Result}

\begin{figure}[t]
  \centering
  \includegraphics[width=8.6cm]{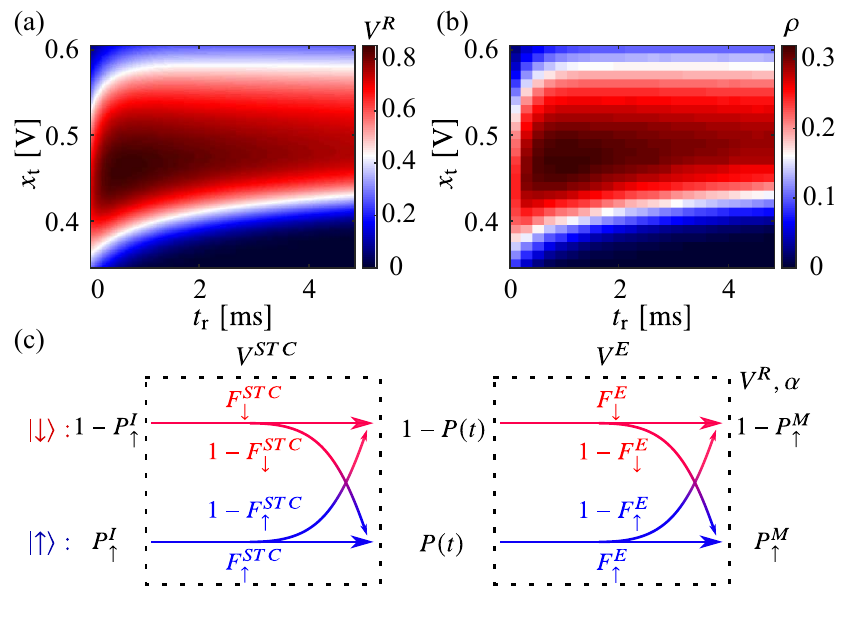}% Here is how to import EPS art
  \caption{\label{fig:3} (a)(b) $V^R$ and the amplitude ($\rho$) as a function 
  of $t_{\rm r}$ and $x_{\rm t}$, \highlight{showing the} consistency in scaled 
  color image. (c) The qubits are prepared in $\ket{\uparrow}$ with 
  \highlight{probability} $P^I_\uparrow$. Throughout the STC conversion 
  ($F_\uparrow^{STC}$, $F_\downarrow^{STC}$), the electron tunnels out from QD 
  with probability $P(t)$. Similarly, the electrons are measured as 
  $\ket{\uparrow}$ with $P_\uparrow^M$ throughout the electrical detection 
  \highlight{process} ($F_\uparrow^{E}$, $F_\downarrow^{E}$).} 
\end{figure}

\subsection{\label{sec:relation}Relation between $P^M_\uparrow$, $P^I_\uparrow$ and $V^R$}
%We now detail the meaning of the readout visibility. During the measurement of decaying time, we fit a decay of 
% the spin-up probability with an exponential of waiting time $t_{wait}$:
% \begin{equation}
%   P_\uparrow = \rho e^{-Wt_{wait}}+\alpha
%   \label{eq:decay time}
% \end{equation}
% where $\rho$ is the amplitude, $W= 1/T_1$ is the spin relaxation rate and $\alpha$ is the dark count. The amplitude
% $\rho$, together with the dark count $\alpha$, characterizes the efficiency of the readout process.

Now, we focus on the details of readout visibility $V^R$. First, we 
\highlight{exhibit the relationship between $V^R$ and readout time $t_{\rm r}$ 
along with threshold voltage $x_{\rm t}$ via the Monte-Carlo method} in 
Fig.~\ref{fig:3}(a). As mentioned in Sec.\ref{sec:readout 
Visibility}, $x_{\rm t}$ \highlight{corresponding to} the maximum 
electrical detection 
visibility $V^E$ increases as $t_{\rm r}$ increases. \highlight{The 
state-to-charge visibility $V^{STC}$ is a uni-modal function of $t_{\rm r}$.} 
Therefore, $V^R$ increases first and then decreases along the $t_{\rm r}$ axis, 
and $x_{\rm t}$ of the maximum $V^R$ increases along the $x_{\rm t}$ axis as 
$t_{\rm r}$ increases.

Then, we drew the spin up probability $\rho$ in the same range in 
Fig.~\ref{fig:3}(b) for comparison. \highlight{The expression} 
$\rho=P^M_\uparrow|_{t_{\rm wait} \to 0}-\alpha$ is obtained by fitting the 
experimental data of the spin relaxation process with an exponential 
\highlight{function} $P^M_\uparrow = \rho\cdot e^{-t/T_1}+\alpha$. Here, 
$\alpha=P^M_\uparrow|_{t_{\rm wait} \to +\infty}$ is the dark count of the spin 
relaxation process. Fig.~\ref{fig:3}(a) and (b) shows that the readout 
visibility $V^R$ and probability $\rho$ are consistent in the \highlight{scaled 
color images}. 

To analyze this consistency, we focus on the details of Eq.\ref{eq:readout visibility}. As shown in Fig.~\ref{fig:3}(c), we note the spin-up probability at the beginning of the readout phase as the prepared probability $P^I_\uparrow$. Throughout the STC conversion, the probability of the tunneling events detected within readout time $t_{\rm r}$ ($P(t)$) depends on the condition probability that electrons in $\ket{\uparrow}$ ($F_\uparrow^{STC}$) or $\ket{\downarrow}$ ($1-F_\downarrow^{STC}$) tunnel out:

\begin{equation}
  P(t) = F_\uparrow^{STC} P^I_\uparrow+(1-F_\downarrow^{STC})(1-P^I_\uparrow).
  \label{eq:probability read}
\end{equation}
Similarly, by comparing the maximum of each readout traces $x_{\rm max}$ with threshold voltage $x_{\rm t}$, the electron is measured with $P^M_\uparrow$ throughout the electrical detection ($F_\uparrow^{E}$, $F_\downarrow^{E}$):
\begin{eqnarray}
  P^M_\uparrow&&=P(t)F_\uparrow^E+(1-P(t))(1-F_\downarrow^E).%\nonumber\\
  % &&=P^I_\uparrow(F_\downarrow^{STC}+F_\uparrow^{STC}-1)(F_\uparrow^E+F_\downarrow^E-1)\nonumber\\
  % +&&F_\uparrow^E-F_\downarrow^{STC}(F_\uparrow^E+F_\downarrow^E-1)\nonumber\\
  %&&=P^I_\uparrow\times V^{STC}V^E+\left(1-F_\downarrow\right)%F_\uparrow^E-F_\downarrow^{STC} V^E
  \label{eq:amplitude plus dark count}
\end{eqnarray}

We \highlight{factorized} Eq.~\ref{eq:amplitude plus dark count} into sectors 
with and without $P^I_\uparrow$ and \highlight{substituted} Eq.~\ref{eq:readout 
fidelity} into Eq.~\ref{eq:amplitude plus dark count} to obtain the expression 
of $\alpha$: $\alpha=1-F_\downarrow^R$. Then, the relation between the measured 
probability $P^M_\uparrow$, prepared probability $P^I_\uparrow$ and readout 
visibility $V^R$ can be \highlight{extracted} by substituting 
Eq.~\ref{eq:readout visibility} into Eq.~\ref{eq:amplitude plus dark count}:
\begin{eqnarray}
  % \alpha=1-F_\downarrow
  P^M_\uparrow=P^I_\uparrow\times V^{R}+\alpha.
  \label{eq: relation}
\end{eqnarray}
The details of derivation are \highlight{discussed} in Sec.~6 of the 
Supplementary Materials.

Eq.~\ref{eq: relation} reveals that the measured probability $P^M_\uparrow$ 
linearly depends on the readout visibility $V^R$ and dark count $\alpha$. Here, 
the slope is the prepared probability $P^I_\uparrow$ and the intercept is the 
dark count. By substituting Eq.~\ref{eq: relation} into the definition of the 
probability $\rho=P^M_\uparrow|_{t_{\rm wait} \to 0}-\alpha$, we 
\highlight{got} $\rho=P^I_\uparrow|_{t_{\rm wait} \to 0}\times V^{R}$. Since 
$P^I_\uparrow$ only depends on the "Wait" process, the probability $\rho$ is 
proportional to the readout visibility $V^R$ in the readout process, and this 
proportional relation between $\rho$ and $V^R$ explains their consistency in 
the \highlight{scaled color images} shown in Fig.~\ref{fig:3}(a) and (b). Since 
$P^I_\uparrow$ only depends on the "Wait" process, we try to apply the 
threshold-independent data process methods in 
Sec.\ref{sec:Threshol-independent}.

\begin{figure}[t]
  \centering 
  \includegraphics[width=8.6cm]{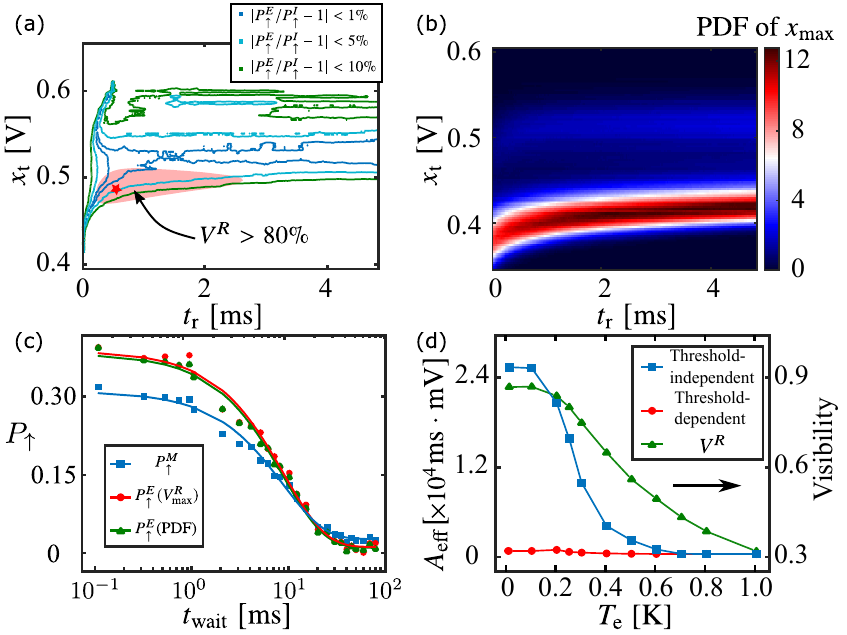}% Here is how to import EPS art
  \caption{\label{fig:4} (a) The bias of the expected probability 
  ($P_\uparrow^E$) relative to $P^I_\uparrow$ ($P_\uparrow^E/P^I_\uparrow-1$) 
  as a function of readout time $t_{\rm r}$ and threshold voltage $x_{\rm t}$. 
  The \highlight{regions between} the blue, cyan and green curves represent 
  $A_{\rm eff}$ where $|P_\uparrow^E/P^I_\uparrow-1|<1\%$, $5\%$ and $10\%$, 
  respectively. For comparison, the red star represents the position of the 
  maximum readout visibility $V^R_{\rm max}$, and the points in the pink shadow 
  region satisfy $V^R>80\%$. (b)\highlight{The} map of PDF of $x_{\rm max}$ 
  (maximum of the readout trace). The monotonically non-decreasing feature of 
  the maximum function causes the positions of the two peaks  to increase as 
  $t_{\rm r}$ increases. The outline of the valley between two peaks is similar 
  to \highlight{that of} (a). (c) The extrapolated probability $P_\uparrow^E$ 
  \highlight{figured out} from the threshold-independent methods at $V^R_{\rm 
  max}$ (red circles) and minimum between two peaks in PDF of $x_{\rm max}$ 
  (green triangles). The measured probabilities $P_\uparrow^M$ (blue 
  rectangles) are obtained from the exponential decay process at $V^R_{\rm 
  max}$ directly. The solid curves with the same color illustrate the 
  corresponding exponential fitting results. $t_{\rm wait}$ is 
  logarithmic, \highlight{showing} that the threshold-independent method 
  suppresses dark count $\alpha$ and improves probability $\rho$. (d) The left 
  y-axis shows \highlight{that $A_{\rm eff}$ is a function} of the electron 
  temperature $T_{\rm e}$ \highlight{in both} threshold-independent method and 
  threshold-dependent readout. The right y-axis shows the corresponding readout 
  visibility $V^R$. When $T_{\rm e}$ exceeds 100 mK, $V^R$ and $A_{\rm eff}$ of 
  the threshold-independent methods \highlight{will decrease} as $T_{\rm e}$ 
  increases. The $A_{\rm eff}$ of the threshold-independent method is higher 
  than that of the threshold-dependent method until $T_{\rm e}=0.7$ K {as $V^R$ 
  fixed at 0.5.}}
\end{figure}

\subsection{\label{sec:Threshol-independent}Threshold-independent data process}

 We calculate the extrapolated probability ($P^E_\uparrow$) for each $t_{\rm r}$ and $x_{\rm t}$ by applying Eq.~\ref{eq: relation} directly:

\begin{eqnarray}
  P^E_\uparrow=(P^M_\uparrow-\alpha)/V^{R}.
  \label{eq: P^E}
\end{eqnarray}
Here, $P^E_\uparrow$ is the extrapolated probability \highlight{and can be 
calculated from} by $V^R$ and $\alpha$. Since $P^I_\uparrow$ is independent of 
the threshold voltage $x_{\rm t}$ and readout time $t_{\rm r}$, we 
\highlight{speculated that} the extrapolated probability $P^E_\uparrow$ 
\highlight{is} threshold independent as well with no constraints on $x_{\rm t}$ 
and $t_{\rm r}$. We calculate the contour map of 
$|P^E_\uparrow/P^I_\uparrow-1|$, as shown in Fig.~\ref{fig:4}(a). The region 
inside the blue, cyan and green curves \highlight{represent} the area where 
$|P_\uparrow^E/P^I_\uparrow-1|<1\%$, $5\%$ and $10\%$, respectively. We 
\highlight{compared} the regions with the visibility $V^R$ map in 
Fig.~\ref{fig:4}(a). The pink shadow represents where $V^R>80\%$, and the red 
star represents the position of $V^R_{\rm max}$. There is a discrepancy between 
$|P^E_\uparrow/P^I_\uparrow-1|$ and the $V^R$ map, \highlight{revealing} that 
the optimal $x_{\rm t}$ and $t_{\rm r}$ for the readout visibility $V^R$ are 
not optimal for $P^E_\uparrow$.

In Sec.5 of the Supplementary Materials, we \highlight{will} use the simulated 
traces to illustrate that the region $|P_\uparrow^E/P^I_\uparrow-1|<1\%$ almost 
covers the whole considered region under ideal circumstances. We 
\highlight{calculated} the cumulative error (CE) and absolute value of error 
between the distribution of experimental and simulated traces as a function of 
threshold voltage $x_{\rm t}$. The shape of the CE curve 
\highlight{demonstrates} that the error in the fitting process comes from the 
bin edges. The features of the bin edge error are shown in the contour map of 
$|P^E_\uparrow/P^I_\uparrow-1|$ in Fig.~2(b) and (c) in the Supplementary 
Materials.

In the valley region between the two peaks, errors from bin edges are minimal, 
as shown in the map of the distribution of $x_{\rm max}$ in 
Fig.~\ref{fig:4}(a). The outline of the valley region in Fig.~\ref{fig:4}(b) 
\highlight{resembles that} of $|P^E_\uparrow/P^I_\uparrow-1|<1\%$ and 5$\%$ in 
Fig.~\ref{fig:4}(a). As a compromise method, we can choose $x_{\rm t}$ around 
the minimum between two peaks of the distribution of $x_{\rm max}$ instead of 
$x_{\rm t}$ at $V^R_{\rm max}$.

By \highlight{using} the threshold-independent method to \highlight{processing} 
the experimental data, we \highlight{calculated} the extrapolated probability 
$P^E_\uparrow$ $t_{\rm r}$ and threshold voltage $x_{\rm t}$ at $V^R_{\rm max}$ 
($P^E_\uparrow(V^R_{\rm max})$), \highlight{along with} the minimum between two 
peaks in PDF of the maximum of each readout trace $x_{\rm max}$ with the same 
$t_{\rm r}$ ($P^E_\uparrow({\rm PDF})$). Fig.~\ref{fig:4}(c) shows 
$P^E_\uparrow$ as a function of waiting time of the spin decaying process 
$t_{\rm wait}$, and we plot measured probability $P^M_\uparrow$ for comparison. 
The results demonstrate that the threshold-independent method suppresses dark 
count $\alpha$ and increases probability $\rho=P^M_\uparrow|_{t_{\rm wait} \to 
0}-\alpha$, and $P^E_\uparrow({\rm PDF})$ calculated via the compromise method 
is slightly different from $P^E_\uparrow(V^R_{\rm max})$. $x_{\rm t}$ and 
readout time $t_{\rm r}$ in a wider range with only a $1\%$ loss of accuracy.

Finally, we \highlight{tried} to use the area where 
$|P^E_\uparrow/P^I_\uparrow-1|<1\%$ as $A_{\rm eff}$ to quantify the accuracy 
and efficiency \highlight{of} different data processing methods. 
For comparison, we use the area of where $|P^M_\uparrow/P^I_\uparrow-1|<1\%$ as $A_{\rm eff}$ for the commonly used threshold-dependent method.
The $A_{\rm eff}$ of the threshold-independent method is 60 times larger than 
that of the threshold-dependent method, \highlight{meaning} that we can choose 
threshold voltage $x_{\rm t}$ and readout time $t_{\rm r}$ in a 60 times 
\highlight{larger} range and maintain 99$\%$ accuracy. In addition, the 
threshold-independent method is more robust and can calibrate the measured 
result from the interference of the experimental hardware limitation.

% We compare $A_{\rm eff}$ in Fig.~\ref{fig:4}(a) with $V^R$, and find the region of $A_{\rm eff}$ can be devided into two parts. The top part has $V^R>70\%$ and the bottom part has $V^R<70\%$. For more common situation, we draw the area where $|P^E_\uparrow/P^I_\uparrow-1|<5\%$ and compare it with the area where $V^R>0.5$. We find the size and shape of two areas are similar, but two areas do not completely overlap.

% To explain this. we substitute Eq.~\ref{eq:readout visibility} and $\alpha=1-F_\downarrow$ into Eq.~\ref{eq: relation}, and obtain the expression of the measured probability of $F_\uparrow$ and $F_\downarrow$:
% \begin{eqnarray}
%   P^M_\uparrow &&= P^I_\uparrow(F_\uparrow+F_\downarrow-1)+(1-F_\downarrow)\nonumber\\
%   &&= P^I_\uparrow F_\uparrow + (P^I_\uparrow-1)(F_\downarrow-1).
%   \label{eq:linear expression}
% \end{eqnarray}
% Here, we define the rest part of $P^M_\uparrow$ as $R=P^M_\uparrow-[P^I_\uparrow F_\uparrow + (P^I_\uparrow-1)(F_\downarrow-1)]$. Fig.~\ref{fig:4}(b) shows $R$ as a function of $F_\uparrow^R$ and $F_\downarrow^R$. The orange dots represent $|R|<1\%$ and the blue dots represent $|R|>1\%$. 

% The successful ratio can be defined as the length of the line segment where the bias of the fitted value is less than 1\% to the total length of the line segment. So in our data process, we fix $t_{\rm r}$ first and use the linear regression of $P^M_\uparrow$, $F_\uparrow $ and $F_\downarrow$ to adjust $P^I_\uparrow$.

\subsection{\label{sec:influence}Influence of the Electron Temperature}

Furthermore, we try to characterize the influence of $T_{\rm e}$. 
Ref.~\cite{keith2019benchmarking} \highlight{assumes} that the tunneling rate 
follows a Fermi distribution:
\begin{eqnarray}
  \Gamma^{\rm out}_{\uparrow,\ \downarrow}=&&[1-f(\epsilon\pm E_z/2,\ T_{\rm e})]\Gamma^{\rm out},\nonumber\\
  \Gamma^{\rm in}_{\uparrow,\ \downarrow}=&&f(\epsilon\pm E_z/2,\ T_{\rm e})\Gamma^{\rm in},
  \label{eq:tunneling rate distribution}
\end{eqnarray}
where $f(\epsilon\pm E_z/2,\ T_{\rm e})$ is the Fermi-Dirac function with $-$ 
for $\ket{\downarrow}$ and $+$ for $\ket{\uparrow}$,\ $\Gamma^{\rm 
out}(\Gamma^{\rm in})$ is the maximum tunnel out(in) rate, $\Gamma^{\rm 
out}_{\uparrow,\ \downarrow}(\Gamma^{\rm in}_{\uparrow,\ \downarrow})$ are 
tunneling rates \highlight{of corresponding} spin state, and $\epsilon$ is the 
energy \highlight{splitting} between the Fermi energy of the electron reservoir 
and the average energy of $\ket{\uparrow}$ and $\ket{\downarrow}$ electrons in 
QD. Therefore, $\epsilon+E_z/2$($\epsilon-E_z/2$) represents the energy 
\highlight{splitting} between the Fermi energy of the electron reservoir and 
the energy state of $\ket{\uparrow}$($\ket{\downarrow}$) electron in QD.

By defining $R_G=\Gamma_{\uparrow}^{\rm out}/\Gamma_{\downarrow}^{\rm out}$, $\epsilon$ can be obtained as follows: 
\begin{eqnarray}
  \epsilon=-k_B T_{\rm e}{\rm ln}\left(\frac{1-R_G}{R_G}\frac{1}{e^{\frac{-E_z}{2k_B T_{\rm e}}}-e^{\frac{E_z}{2k_B T_{\rm e}}}}\right).
  \label{eq:epsilon}
\end{eqnarray}

We \hl[extracted] the maximum tunneling rates $\Gamma^{\rm out}$ and $\Gamma^{\rm in}$ by substituting $\epsilon$ into Eq.~\ref{eq:tunneling rate distribution}.

Then, we \highlight{tried} to simulate the single-shot readout process 
\highlight{at} different electron temperature $T_{\rm e}$. Assuming that 
tunneling rates $\Gamma^{\rm out}$ and $\Gamma^{\rm in}$ are not associated 
with $T_{\rm e}$, we directly \highlight{substituted} $T_{\rm e}$ into 
Eq.~\ref{eq:tunneling rate distribution} to calculate the tunneling rates. We 
\highlight{used} the Monte-Carlo method to generate the simulated traces with a 
fixed external magnetic field $B_{\rm ext}=1.5$ T \highlight{at} different 
$T_{\rm e}$. The left y-axis in Fig.~\ref{fig:4}(d) shows the $A_{\rm eff}$ of 
\hl[both] the threshold-independent method and the threshold-dependent method 
at different $T_{\rm e}$. The right y-axis shows the corresponding readout 
visibility $V^R$. The simulation results \highlight{exhibit} that the $A_{\rm 
eff}$ of the threshold-independent methods is 60 times greater than that of the 
threshold-dependent method when $T_{\rm e}<0.1$ K. As $T_{\rm e}$ increases, 
$A_{\rm eff}$ of the threshold-independent methods decreases. It is larger than 
\highlight{that of} the threshold-dependent method until $T_{\rm e}=0.7$ K. 
When $T_{\rm e}>0.7$ K, the corresponding $V^R<0.5$. Here, we \hl[can] give the 
boundary condition of the threshold-independent method as $T_{\rm e}=0.7$ K 
when $B_{\rm ext}=1.5$ T.

\section{\label{sec:Conclusion}Summary}
% The single-shot readout data process will become increasingly important as the experiment progresses to the fault tolerance threshold of the logic qubit~\cite{huang2019two-qubit, Vandersypen2020hot-qubit, Tarucha2021three-qubit}. $V^R$ has become the limiting source of infidelity, and the current research focuses on the improvements of the visibilities. 
We \highlight{described} a threshold-independent method \highlight{of} the 
single-shot readout data process based on the linear dependence of measured 
probability $P_\uparrow^M$ with the corresponding readout visibility $V^R$ and 
dark count $\alpha$. Due to the error during the fitting process from bin 
edges, the extrapolated probability deviates from the prepared probability. 
\highlight{For compromise}, the region of readout time $t_{\rm r}$ and 
threshold voltage $x_{\rm t}$ \highlight{are reduced to} the minimum of the 
distribution of the maximum of each readout trace $x_{\rm max}$ to ensure that 
the accuracy loss of extrapolated probability $P_\uparrow^E$ is less than 
$1\%$. Then, we \highlight{used} $A_{\rm eff}$ to quantify the efficiency of 
the threshold-independent method and the threshold-dependent readout. The 
result shows that the $A_{\rm eff}$ of the threshold-independent method is more 
than 60 times larger \highlight{that of the threshold-dependent method}. 
Moreover, we \highlight{simulated} the single-shot readout process at different 
electron temperature $T_{\rm e}$. We broaden the boundary condition of the 
single-shot readout to 0.7 K with $B_{\rm ext}=1.5$ T, where $V^R=0.5$. 
\highlight{The significance of employing the threshold-independent method will 
progressively increase as the experiment advances towards} the fault tolerance 
threshold of the logic qubit, \highlight{particularly when operating under high 
electron temperature conditions}~\cite{huang2019two-qubit, 
Vandersypen2020hot-qubit, Tarucha2021three-qubit}.

\addcontentsline{toc}{chapter}{Acknowledgment}
\section*{Acknowledgment}
This work was supported by the National Natural Science Foundation of China (Grants No. 12074368, 92165207, 12034018 and 62004185), the Anhui Province Natural Science Foundation (Grants No. 2108085J03), the USTC Tang Scholarship, and this work was partially carried out at the USTC Center for Micro and Nanoscale Research and Fabrication.


\begin{thebibliography}{999}

\bibitem{Muhonen2014Storing}
Juha~T. Muhonen, Juan~P. Dehollain, Arne Laucht, et~al.
\newblock Storing quantum information for 30 seconds in a nanoelectronic
  device.
\newblock {\em Nature Nanotechnology}, 9:986, 2014.

\bibitem{Zhang2018Semiconductor}
Xin Zhang, Hai-Ou Li, Gang Cao, et~al.
\newblock Semiconductor quantum computation.
\newblock {\em National Science Review}, 6:32, 2019.
  
\bibitem{Eriksson2020fabtication}
J.~P. Dodson, Nathan Holman, Brandur Thorgrimsson, et~al.
\newblock Fabrication process and failure analysis for robust quantum dots in
  silicon.
\newblock {\em Nanotechnology}, 31:505001, 2020.
  
\bibitem{Veldhorst2018scalable}
Ruoyu Li, Luca Petit, David~P. Franke, et~al.
\newblock A crossbar network for silicon quantum dot qubits.
\newblock {\em Science Advances}, 4:eaar3960, 2018.

\bibitem{Camenzind2021Spin}
Leon~C Camenzind, Simon Geyer, Andreas Fuhrer, et~al.
\newblock A spin qubit in a fin field-effect transistor.
\newblock {\em arXiv preprint}, 2103:07369, 2021.

\bibitem{Zwerver2022Qubits}
A.~M.~J. Zwerver, T.~Krähenmann, T.~F. Watson, et~al.
\newblock Qubits made by advanced semiconductor manufacturing.
\newblock {\em Nature Electronics}, 5:184, 2022.
  
\bibitem{Tarucha2018single-qubit}
J.~Yoneda, K.~Takeda, T.~Otsuka, et~al.
\newblock A quantum-dot spin qubit with coherence limited by charge noise and
  fidelity higher than 99.9\%.
\newblock {\em Nature Nanotechnology}, 13:102, 2018.

\bibitem{Dzurak2018single-qubit}
K.~W. Chan, W.~Huang, C.~H. Yang, et~al.
\newblock Assessment of a silicon quantum dot spin qubit environment via noise
  spectroscopy.
\newblock {\em Phys. Rev. Applied}, 10:044017, 2018.
  
\bibitem{Xue2022Qunatum}
Xiao Xue, Maximilian Russ, Nodar Samkharadze, et~al.
\newblock Quantum logic with spin qubits crossing the surface code threshold.
\newblock {\em Nature}, 601:343, 2022.
  
\bibitem{Noiri2022Fast}
Akito Noiri, Kenta Takeda, Takashi Nakajima, et~al.
\newblock Fast universal quantum gate above the fault-tolerance threshold in
  silicon.
\newblock {\em Nature}, 601:338, 2022.

\bibitem{Mills2022Tow-qubit}
Adam~R. Mills, Charles~R. Guinn, Michael~J. Gullans, et~al.
\newblock Two-qubit silicon quantum processor with operation fidelity exceeding
  99\%.
\newblock {\em Science Advances}, 8:eabn5130, 2022.

\bibitem{elzerman2004single}
JM~Elzerman, R~Hanson, LH~Willems van Beveren, et~al.
\newblock Single-shot read-out of an individual electron spin in a quantum dot.
\newblock {\em Nature}, 430:431, 2004.

\bibitem{Zajac2016Scable}
D.~M Zajac, T.~M Hazard, X.~Mi, E.~Nielsen, and J.~R Petta.
\newblock Scalable gate architecture for a one-dimensional array of
  semiconductor spin qubits.
\newblock {\em Physical Review Applied}, 6(5), 2016.


% \bibitem{Petersson2010Rf_readout}
% K.~D. Petersson, C.~G. Smith, D.~Anderson, et~al.
% \newblock Charge and spin state readout of a double quantum dot coupled to a
%   resonator.
% \newblock {\em Nano Letters}, 10:2789, 2010.
  
\bibitem{Pakkiam2018single-shot}
P.~Pakkiam, A.~V. Timofeev, M.~G. House, et~al.
\newblock Single-shot single-gate rf spin readout in silicon.
\newblock {\em Phys. Rev. X}, 8:041032, 2018.

\bibitem{West2019Gate-based}
Anderson West, Bas Hensen, Alexis Jouan, et~al.
\newblock Gate-based single-shot readout of spins in silicon.
\newblock {\em Nature Nanotechnology}, 14:437, 2019.

\bibitem{Zheng2019RapidReadout}
Guoji Zheng, Nodar Samkharadze, Marc~L. Noordam, et~al.
\newblock Rapid gate-based spin read-out in silicon using an on-chip resonator.
\newblock {\em Nature Nanotechnology}, 14:742, 2019.
  
\bibitem{Prance2015identifying}
J.~R. Prance, B.~J.~Van Bael, C.~B. Simmons, et~al.
\newblock Identifying single electron charge sensor events using wavelet edge
  detection.
\newblock {\em Nanotechnology}, 26:215201, 2015.
  
\bibitem{Nowack2011single-shot}
K.~C. Nowack, M.~Shafiei, M.~Laforest, et~al.
\newblock Single-shot correlations and two-qubit gate of solid-state spins.
\newblock {\em Science}, 333:1269, 2011.

\bibitem{anjou2014optimal}
B.~D'Anjou and W.~A. Coish.
\newblock Optimal post-processing for a generic single-shot qubit readout.
\newblock {\em Phys. Rev. A}, 89:012313, 2014.

% \bibitem{Bagrets2003full}
% D.~A. Bagrets and Yu.~V. Nazarov.
% \newblock Full counting statistics of charge transfer in coulomb blockade
%   systems.
% \newblock {\em Phys. Rev. B}, 67:085316, 2003.

% \bibitem{Contreras2014dephasing}
% L~D Contreras-Pulido, M~Bruderer, S~F Huelga, and M~B Plenio.
% \newblock Dephasing-assisted transport in linear triple quantum dots.
% \newblock {\em New Journal of Physics}, 16:113061, 2014.

\bibitem{Struck2021Neural}
Tom Struck, Javed Lindner, Arne Hollmann, et~al.
\newblock Robust and fast post-processing of single-shot spin qubit detection
  events with a neural network.
\newblock {\em Scientific Reports}, 11(1):16203, 2021.


\bibitem{gorman2017tunneling}
S.~K. Gorman, Y.~He, M.~G. House, et~al.
\newblock Tunneling statistics for analysis of spin-readout fidelity.
\newblock {\em Phys. Rev. Applied}, 8:034019, 2017.

\bibitem{Mizokuchi2020Detection}
Raisei Mizokuchi, Masahiro Tadokoro, and Tetsuo Kodera.
\newblock Detection of tunneling events in physically defined silicon quantum
  dot using single-shot measurements improved by numerical treatments.
\newblock {\em Applied Physics Express}, 13(12):121004, nov 2020.
    
\bibitem{merello2010single}
Andrea Morello, Jarryd~J. Pla, Floris~A. Zwanenburg, et~al.
\newblock Single-shot readout of an electron spin in silicon.
\newblock {\em Nature}, 467:687, 2010.
  
\bibitem{veldhorst2014addressable}
M.~Veldhorst, J.~C.~C. Hwang, C.~H. Yang, et~al.
\newblock An addressable quantum dot qubit with fault-tolerant
  control-fidelity.
\newblock {\em Nature Nanotechnology}, 9:981, 2014.
  
\bibitem{Kawakami2014electrical}
E.~Kawakami, P.~Scarlino, D.~R. Ward, et~al.
\newblock Electrical control of a long-lived spin qubit in a {Si/SiGe} quantum
  dot.
\newblock {\em Nature Nanotechnology}, 9:666, 2014.

\bibitem{Vukusic2018single-shot}
Lada Vukušić, Josip Kukučka, Hannes Watzinger, et~al.
\newblock Single-shot readout of hole spins in {Ge}.
\newblock {\em Nano Letters}, 18:7141, 2018.

\bibitem{Buch2013spin}
H.~Büch, S.~Mahapatra, R.~Rahman, et~al.
\newblock Spin readout and addressability of phosphorus-donor clusters in
  silicon.
\newblock {\em Nature Communications}, 4:2017, 2013.

\bibitem{robledo2011high-fidelity}
Lucio Robledo, Lilian Childress, Hannes Bernien, et~al.
\newblock High-fidelity projective read-out of a solid-state spin quantum
  register.
\newblock {\em Nature}, 477:574, 2011.

\bibitem{Mills2022High_Fidelity}
A.~R. Mills, C.~R. Guinn, M.~M. Feldman, et~al.
\newblock High fidelity state preparation, quantum control, and readout of an
  isotopicallyenriched silicon spin qubit.
\newblock {\em arXiv preprint}, 2204:09551, 2022.

\bibitem{keith2019benchmarking}
D~Keith, SK~Gorman, L~Kranz, et~al.
\newblock Benchmarking high fidelity single-shot readout of semiconductor
  qubits.
\newblock {\em New Journal of Physics}, 21:063011, 2019.

\bibitem{xiao2010measurement}
Ming Xiao, MG~House, and Hong~Wen Jiang.
\newblock Measurement of the spin relaxation time of single electrons in a
  silicon metal-oxide-semiconductor-based quantum dot.
\newblock {\em Phys. Rev. Lett.}, 104:096801, 2010.

\bibitem{Carroll2013electron}
L.~A. Tracy, T.~M. Lu, N.~C. Bishop, et~al.
\newblock Electron spin lifetime of a single antimony donor in silicon.
\newblock {\em Applied Physics Letters}, 103:143115, 2013.
  
\bibitem{house2013detection}
M.~G. House, Ming Xiao, GuoPing Guo, et~al.
\newblock Detection and measurement of spin-dependent dynamics in random
  telegraph signals.
\newblock {\em Phys. Rev. Lett.}, 111:126803, 2013.

\bibitem{Zhang2020Giant}
Xin Zhang, Rui-Zi Hu, Hai-Ou Li, et~al.
\newblock Giant anisotropy of spin relaxation and spin-valley mixing in a
  silicon quantum dot.
\newblock {\em Phys. Rev. Lett.}, 124:257701, 2020.

\bibitem{Hu2021Operation}
Rui-Zi Hu, Rong-Long Ma, Ming Ni, et~al.
\newblock An operation guide of {Si-MOS} quantum dots for spin qubits.
\newblock {\em Nanomaterials}, 11:2486, 2021.

\bibitem{huang2019two-qubit}
W.~Huang, C.~H. Yang, K.~W. Chan, et~al.
\newblock Fidelity benchmarks for two-qubit gates in silicon.
\newblock {\em Nature}, 569:532, 2019.

\bibitem{Vandersypen2020hot-qubit}
L.~Petit, H.~G.~J. Eenink, M.~Russ, et~al.
\newblock Universal quantum logic in hot silicon qubits.
\newblock {\em Nature}, 580:355, 2020.

\bibitem{Tarucha2021three-qubit}
Kenta Takeda, Akito Noiri, Takashi Nakajima, et~al.
\newblock Quantum tomography of an entangled three-qubit state in silicon.
\newblock {\em Nature Nanotechnology}, 16:965, 2021.

\end{thebibliography}
\end{document}

\title{Supplementary Material: Threshold-independent method for single-shot readout of spin qubits in semiconductor quantum dots}

\author{Rui-Zi Hu$^{1,2}$, Sheng-Kai Zhu$^{1,2}$, Xin Zhang$^{1,2}$, Yuan Zhou$^{1,2}$, Ming Ni$^{1,2}$, \\ Rong-Long Ma$^{1,2}$, Gang Luo$^{1,2}$, Zhen-Zhen Kong$^{3}$, Gui-Lei Wang$^{3,4}$, Gang Cao$^{1,2}$, \\ Hai-Ou Li$^{1,2}$\thanks{E-mail: haiouli@ustc.edu.cn}, Guo-Ping Guo$^{1,2,5}$\\
$^{1}${CAS Key Laboratory of Quantum Information,} \\ {University of Science and Technology of China, Hefei, Anhui 230026, China}\\  
$^{2}${CAS Center For Excellence in Quantum Information and Quantum Physics,} \\ {University of Science and Technology of China, Hefei 230026, China}\\ 
$^{3}${Key Laboratory of Microelectronics Devices and Integrated Technology,} \\{Institute of Microelectronics, Chinese Academy of Sciences, Beijing 100029, China}\\
$^{4}${Beijing Superstring Academy of Memory Technology, Beijing 100176, China}\\
$^{5}${Origin Quantum Computing Company Limited, Hefei, Anhui 230026, China}} 

\date{\today}
\maketitle

\section{\label{sec:simulation}Simulation of the single-shot readout process}
\begin{figure}[t]
    \centering 
    \includegraphics[width=8.6cm]{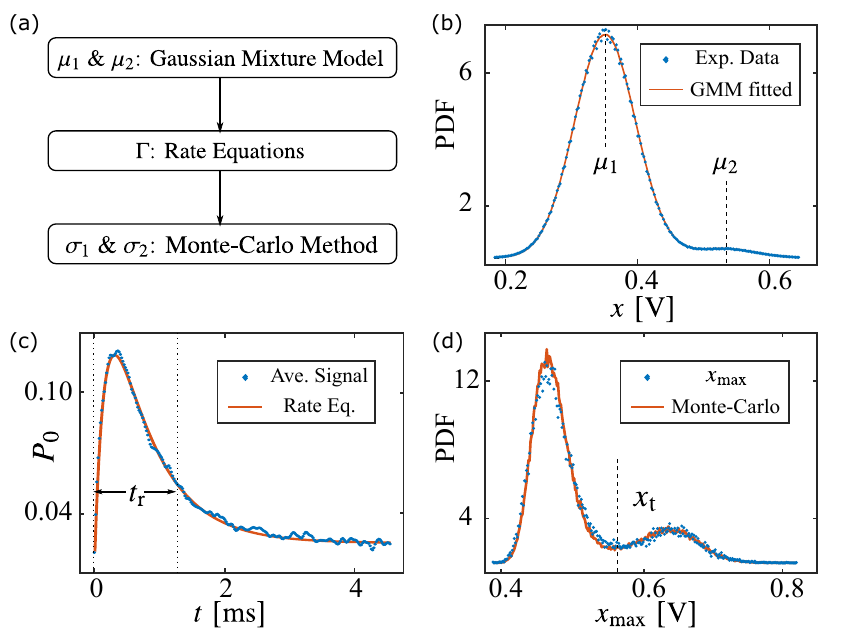}% Here is how to import EPS art
    \caption{\label{fig:5} (a) \highlight{A three-step process} are applied to 
    extract the parameters of the single-shot readout process in order: (b) We 
    use the Gaussian Mixture Model (GMM) method to fit the PDF of data in the 
    readout phase and obtain the mean value of the traces of the occupied and 
    ionized QDs $\mu_1$ and $\mu_2$. Fig.~1(c) in the main text shows two 
    typical readout traces\highlight{selected} among all 1000 repeated 
    measurements; (c) $\Gamma^{\rm in,out}$ are fitted from $\bar{x}$ via the 
    rate equations. $\bar{x}$ is normalized by $\mu_1$ and $\mu_2$. (d) We 
    generate the simulated readout traces without noise via the Monte-Carlo 
    method by using $\mu_1$, $\mu_2$ and $\Gamma^{\rm in,out}$. The standard 
    deviations of the readout traces ($\sigma_1$, $\sigma_2$) are fitted from 
    the PDF of $x_{\rm max}$. The dashed line represents a typical position of 
    $x_{\rm t}$.}
\end{figure}

We use \highlight{a three-step process} to fit the parameters of the 
single-shot readout traces, as shown in Fig.~\ref{fig:5}(a). First, 
\highlight{by fitting the distribution of all data in the readout phase of 
readout traces $x$ with the Gaussian Mixture Model (GMM) method, we can 
extract} the average value of the traces of the occupied (ionized) QD, $\mu_1$ 
($\mu_2$), as shown in Fig.~\ref{fig:5}(b). Then, we normalize the averaged 
readout traces ($\bar{x}$) by $\mu_1$ and $\mu_2$ and fit $\Gamma^{\rm in,out}$ 
by using the rate equations shown in Fig.~\ref{fig:5}(c). Finally, with the 
knowledge of $\Gamma^{\rm in,out}$ and $\mu_{1,2}$, we generate the simulated 
readout traces without noise via the Monte-Carlo method. After adding the noise 
to the simulated traces, we fit the standard deviations ($\sigma_1$, 
$\sigma_2$) from the distribution of \added{maximum of each readout trace 
}$x_{\rm max}$ in Fig.~\ref{fig:5}(d). Since the Monte-Carlo method is used 
only once to generate the traces, it reduces the running time of the fitting 
process while ensuring \highlight{the} accuracy.

\section{\label{sec:S.V.}The State-to-Charge Visibility}

We employed a rate equation model to determine the expectation value of the 
number of electrons on QDs~\cite{xiao2010measurement}. The probabilities of the 
electron in three states are contained in the vector 
$\mathbf{P}=(P_\uparrow,P_\downarrow,P_0)$. The rate equation 
$d\mathbf{P}/dt=\mathbf{QP}$ describes the evolution of $\mathbf{P}$ 
\highlight{over} time. Here, $\mathbf{Q}$ is the transition matrix with 
$\Gamma_\uparrow^{\rm in}$ neglected:
\begin{eqnarray}
  \mathbf{Q} =\left(
    \begin{array}{ccc}
        -W-\Gamma_\uparrow^{\rm out}&0&0 \\
        W&-\Gamma_\downarrow^{\rm out}&\Gamma_\downarrow^{\rm in} \\
        \Gamma_\uparrow^{\rm out}&\Gamma_\downarrow^{\rm out}&-\Gamma_\downarrow^{\rm in}
    \end{array}
    \right)
  \label{eq:transition matrix}
\end{eqnarray}

We assume that the electron \highlight{tunnels} into QD in the "Load \& Wait" 
phase, so we obtain $\mathbf{P}(0)=(P_{\uparrow}^i,1-P_{\uparrow}^i,0)$. By 
solving the rate equation, the average \added{readout} trace $\bar{x}(t)$ over 
many periods of pulse sequences reads~\cite{Carroll2013electron}:
\begin{eqnarray}
  P_0(t) &&= (\Gamma_{\downarrow}^{\rm out}/\Gamma_{\downarrow}^{\rm total})(1-e^{-t\Gamma_{\downarrow}^{\rm  total}})\nonumber\\
  &&+P_{\uparrow}^i(\Gamma_{\uparrow}^{\rm out}-\Gamma_{\downarrow}^{\rm out})/(W+\Gamma_{\uparrow}^{\rm out}-\Gamma_{\downarrow}^{\rm  total})\nonumber\\
  &&\times(e^{-t\Gamma_{\downarrow}^{\rm  total}}-e^{-t(W+\Gamma_{\uparrow}^{\rm out})}),
  \label{eq:ave signal}
\end{eqnarray}
where $P_0(t)=(\bar{x}(t)-\mu_1)/(\mu_2-\mu_1)$ is the normalized averaged 
readout traces by $\mu_1$ and $\mu_2$. $P_{\uparrow}^i$ is the probability of a 
spin-up electron occupying QD at the beginning of the readout phase, 
$\Gamma_{\uparrow}^{\rm out}(\Gamma_{\downarrow}^{\rm out})$ is the unloading 
rate for electron spin up (down), $\Gamma_{\downarrow}^{\rm in}$ is the loading 
rate for electron spin down, and $\Gamma_{\downarrow}^{\rm  
total}=\Gamma_{\downarrow}^{\rm out}+\Gamma_{\downarrow}^{\rm in}$ 
\highlight{is the total loading rate of electron}.

We ignore the spin relaxation rate $W=1/T_1$ while fitting because $\Gamma_{\uparrow}^{\rm out}\gg  W$ here. Fig.~\ref{fig:5}(c) shows the fitting results of $P_0$ as a function of readout time.

With the tunneling rates fitted, Ref.~\cite{Buch2013spin,keith2019benchmarking} 
\highlight{calculated the corresponding STC conversion fidelities 
($F_\downarrow^{STC}(t),\ F_\uparrow^{STC}(t)$) and $V^{STC}(t)$, and all of 
them are functions of $t_{\rm r}$}:
\begin{eqnarray}
  F_\downarrow^{STC}(t) &&= e^{-t\Gamma_{\downarrow}^{\rm out}},\nonumber\\
  F_\uparrow^{STC}(t) &&= 1-\frac{We^{-t\Gamma_{\downarrow}^{\rm out}}+(\Gamma_{\uparrow}^{\rm out}-\Gamma_{\downarrow}^{\rm out})
  e^{-t(W+\Gamma_{\uparrow}^{\rm out})}}{W+\Gamma_{\uparrow}^{\rm out}-\Gamma_{\downarrow}^{\rm out}},\nonumber\\
  V^{STC}(t) &&= F_\downarrow^{STC}(t)+F_\uparrow^{STC}(t) - 1\nonumber\\
  &&= \frac{\Gamma_{\uparrow}^{\rm out}-\Gamma_{\downarrow}^{\rm out}}{W+\Gamma_{\uparrow}^{\rm out}-\Gamma_{\downarrow}^{\rm out}}
  \left(e^{-t\Gamma_{\downarrow}^{\rm out}}-e^{-t(W+\Gamma_{\uparrow}^{\rm out})}\right).
  \label{eq:stc visibilities}
\end{eqnarray}
The corresponding $t_{\rm opt}^{STC}$ is \highlight{shown in} the following 
\highlight{equation}:
\begin{eqnarray}
  t_{\rm opt}^{STC} = \frac{1}{W+\Gamma_{\uparrow}^{\rm out}-\Gamma_{\downarrow}^{\rm out}}{\rm ln}\left(\frac{W+\Gamma_{\uparrow}^{\rm out}}{\Gamma_{\downarrow}^{\rm out}}\right).
  \label{eq:t_read optimum}
\end{eqnarray}

\section{\label{sec:E.D.V.}Electrical Detection Visibility}
Several methods have been used to extract \added{electrical detection visibility }$V_{E}$, including the analytical expression of the distribution~\cite{Nowack2011single-shot,anjou2014optimal}, statistical techniques~\cite{Bagrets2003full,Contreras2014dephasing,gorman2017tunneling}, and the simulation of the readout traces~\cite{merello2010single}. Theoretically, the probability density $N_{\downarrow,\uparrow}^E(x) $ of $x$ can be resolved well and the electrical detection fidelities can be extracted as follows:
\begin{equation}
  F_i^E=\int_{x_s}^{x_{\rm f}}N_i^E(x)\mathrm{d}x,
  \label{eq:electrical fidelity}
\end{equation}
where the integral bounds are $x_s=-\infty(x_{\rm t})$ and $x_{\rm f}=x_{\rm t}(+\infty)$ for $i=\downarrow(\uparrow)$, where $x_{\rm t}$ is the threshold voltage. $V^{E}$ is defined as:
\begin{eqnarray}
  V^E = F_{\downarrow}^E+F_{\uparrow}^E-1.
  \label{eq:electrical visibility}
\end{eqnarray}
The numerical solution of $\mathrm{d}V^E/\mathrm{d}x = 0$ gives the 
\highlight{optimal} $x_{\rm t}$ where $N_\downarrow(x_{\rm opt}) = 
N_\uparrow(x_{\rm opt})$. However, it is difficult to extract $V^E$ directly in 
practice \highlight{for lacking} of a simple analytical expression for the 
distribution of $x_{\rm max}$~\cite{anjou2014optimal}. 

\section{\label{sec:R.V.}The Readout Visibility}

We use the Monte-Carlo method to extract readout visibility $V^R$. We generate the simulated traces of different spin states using the fitted parameters in Eq~\ref{sec:simulation}. Then, the probability density $N_{\downarrow,\uparrow}(x) $ of $x$ can be calculated from the simulated traces as shown in Fig.~\ref{fig:5}, and the readout fidelities can be extracted directly:
\begin{equation}
  F_i^R=\int_{x_s}^{x_{\rm f}}N_i(x)\mathrm{d}x,
  \label{eq:analytical readout fidelity}
\end{equation}
where the integral bounds are $x_s=-\infty(x_{\rm t})$ and $x_{\rm f}=x_{\rm 
t}(+\infty)$ for $i=\downarrow(\uparrow)$, where $x_{\rm t}$ is the threshold 
voltage. $V^R=F^R_\uparrow+F^R_\downarrow-1$ can be calculated as mentioned in 
the main text. Here, we notice that Eq.~\ref{eq:analytical readout fidelity} 
has \highlight{no correlation} with spin up probability $P_\uparrow$, so 
readout visibility $V^R$ only depends on the in-variants of the measurement, 
e.g., $\Gamma$, $\mu$ and $\sigma$.

\section{\label{Cali of Step}Calibration of the Edges of Bins}

\begin{figure}[t]
    \centering 
    \includegraphics[width=8.6cm]{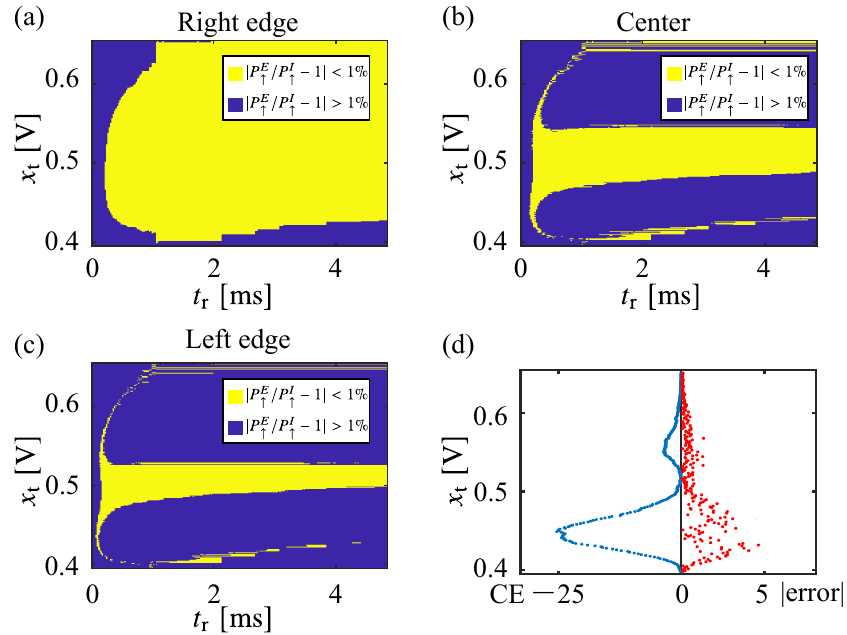}% Here is how to import EPS art
    \caption{\label{fig:6} After calculating $V^R$, an array of bins is 
    obtained. (a) We use the right edge of each bin to calculate 
    $P^M_\uparrow$. The region where $|P^E_\uparrow/P^I_\uparrow-1|<1\%$ almost 
    covers the whole considered region. (b) We use the center of each bin to 
    calculate $P^M_\uparrow$. The interval between the center and left edge is 
    only 1 mV, but the region where $|P^E_\uparrow/P^I_\uparrow-1|<1\%$ 
    reduces. (c) We use the left edge of each bin to calculate $P^M_\uparrow$. 
    The region where $|P^E_\uparrow/P^I_\uparrow-1|<1\%$ reduces again. (d) The 
    blue points in the left row represent the cumulative error (CE) of the 
    distribution between the experimental and simulated traces \highlight{which 
    are all functions} of $x_{\rm t}$; the red points in the right row 
    represent the corresponding absolute value of error ($|{\rm error}|$).}
\end{figure}

Since $x_{\rm t}$ is discrete when extracting \added{readout visibility}$V^R$ in practice, we can obtain an array of bins that contain the left edges, centers and right edges with intervals $\approx1$ mV. We calculate \added{measured probability }$P^M_\uparrow$ and the corresponding \added{extrapolated probability }$P^E_\uparrow=(P^M_\uparrow-\alpha)/V^R$ with the same simulated traces. The regions where $|P^E_\uparrow/P^I_\uparrow-1|<1\%$ are shown in Fig.~\ref{fig:6} by using the right edge(a), center(b) and left edge(c) of each bin of \added{threshold voltage }$x_{\rm t}$. For $P^M_\uparrow$ calculated with the right edges, the region where $|P^E_\uparrow/P^I_\uparrow-1|<1\%$ almost covers the whole considered region as we expect. However, for \added{probability }$P^M_\uparrow$ calculated with the left edges and centers, the dislocation of the edges of bins for \added{visibility }$V^R$ and \added{probability }$P^M_\uparrow$ causes the region where $|P^E_\uparrow/P^I_\uparrow-1|<1\%$ \replaced{to reduce}{reduces}, and the required \added{visibility }$V^R$ increases from $<10\%$ to 70\%. The reduction in area where $|P^E_\uparrow/P^I_\uparrow-1|<1\%$ reveals that the edges of bins should be calibrated carefully before calculating $P^E_\uparrow$.

Fig.~\ref{fig:6}(d) shows the cumulative error (CE) and the absolute value of 
error between the distribution of experimental and simulated traces 
\highlight{which are all functions of} threshold voltage $x_{\rm t}$. The shape 
of the CE curve is similar to the distribution of maximum of each readout trace 
$x_{\rm max}$. We believe that this error comes from the bin edge mismatch 
during the fitting process. 
We carefully check the simulation and fitting process and try to rewrite the fitted function to reduce the bin edge errors.

\section{\label{relation}Relation between $P^M_\uparrow$ and $P^I_\uparrow$}

In Sec. 2.3 of the main text, we give the relation between measured probability 
$P^M_\uparrow$ and prepared probability $P^I_\uparrow$ by 
\highlight{substituting} Eq. 1, 2 and 3 into Eq. 4 in the main text. In this 
section, we will \highlight{describe the derivation process in detail}.

First, we factorize Eq. 3 in the main text into sectors with and without $P^I_\uparrow$:

\begin{eqnarray}
  P(t) &&= F_\uparrow^{STC} P^I_\uparrow+(1-F_\downarrow^{STC})(1-P^I_\uparrow)\nonumber\\
  &&= (F_\uparrow^{STC}+F_\downarrow^{STC}-1)P^I_\uparrow + (1-F_\downarrow^{STC})\nonumber\\
  &&=V^{STC}P^I_\uparrow + (1-F_\downarrow^{STC}).
  \label{eq:Pt}
\end{eqnarray}

Here, we use the definition of state-to-charge visibility $V^{STC}=F_\uparrow^{STC}+F_\downarrow^{STC}-1$ at the last line.

Then, we substitute Eq.~\ref{eq:Pt} into Eq. 4 in the main text and factorize Eq. 4 in the main text into sectors with and without $P^I_\uparrow$:

\begin{eqnarray}
  P^M_\uparrow&&=P(t)F_\uparrow^E+(1-P(t))(1-F_\downarrow^E)\nonumber\\
  &&=V^E P(t)+(1-F_\downarrow^E)\nonumber\\
  &&=V^E V^{STC}P^I_\uparrow%\nonumber\\
  +F_\uparrow^E-F_\downarrow^{STC}(F_\uparrow^E+F_\downarrow^E-1)\nonumber\\
  &&=P^I_\uparrow\times V^R+\left(1-F_\downarrow^R\right)\nonumber\\
  &&=P^I_\uparrow\times V^R+\alpha.%F_\uparrow^E-F_\downarrow^{STC} V^E
  \label{eq:Pm}
\end{eqnarray}
\added{
  Here, we use the relation between electrical detection visibility $V^E$, state-to-charge visibility $V^{STC}$ and readout visibility $V^R$, and the definition of readout fidelity of spin down $F_\downarrow^R$ mentioned in Eq.~1 and 2 in the main text for simplification.
}